\documentclass[conference]{IEEEtran}
\IEEEoverridecommandlockouts
\usepackage{cite}
\usepackage{amsmath,amssymb,amsfonts}
\usepackage{algorithmic}
\usepackage{graphicx}
\usepackage{textcomp}
\usepackage{xcolor}
\usepackage{url}
\usepackage{caption}
\usepackage{hyperref}
\usepackage{booktabs}
\usepackage{multirow}
\usepackage{array}
\hypersetup{%
    pdfborder = {0 0 0}
}
\DeclareCaptionFormat{lowercase}{\MakeLowercase{#1#2#3}}
\def\BibTeX{{\rm B\kern-.05em{\sc i\kern-.025em b}\kern-.08em
    T\kern-.1667em\lower.7ex\hbox{E}\kern-.125emX}}
\begin{document}

\title{Assistant or Actor? Student Trust, Control, and Delegation Regret When Using a General-Purpose AI Agent}

\author{
\IEEEauthorblockN{Shiva Pochampally}
\IEEEauthorblockA{
Department of Computer Science\\
Virginia Tech\\
Blacksburg, VA, USA\\
shivapochampally@vt.edu
}

\and

\IEEEauthorblockN{Shengwei An}
\IEEEauthorblockA{
Department of Computer Science\\
Virginia Tech\\
Blacksburg, VA, USA\\
swan@vt.edu
}

\and

\IEEEauthorblockN{Yan Chen}
\IEEEauthorblockA{
Department of Computer Science\\
Virginia Tech\\
Blacksburg, VA, USA\\
ych@vt.edu
}
}

\maketitle

\begin{abstract}
When AI agents shift from answering questions to taking actions, users face a new problem: deciding what to delegate, to a system whose action space they cannot fully anticipate. We call the resulting dissatisfaction \textit{delegation regret}, a pattern in which users regret not that the agent made an error, but that it acted beyond what they would have authorized. In a controlled study, 20 university students completed five common daily tasks using OpenClaw, a general-purpose AI agent, across tasks chosen to vary in privacy, stakes, and reversibility. For each task we measured perceived success, trust, supervision demand, usefulness, transparency, approval preference, and verification need on 5-point Likert scales, and collected free-text reflections analyzed through thematic coding. Three findings emerged. First, participants calibrated trust per task rather than per agent: they granted wide autonomy for advisory and low-stakes tasks but demanded confirmation for irreversible, externally visible actions. Second, irreversibility combined with external visibility, rather than stakes alone, appeared to drive trust withdrawal: the moderate-stakes email task triggered the sharpest drop in trust ($M = 3.10$) and the highest demand for approval ($M = 4.65$), whereas a high-stakes but verifiable task did not produce the same response.
Third, delegation regret appeared consistently when the agent executed actions without preview, even when the output was rated as successful. 
We discuss implications for agent designs that expose action boundaries, support per-task autonomy policies, and separate advisory output from agentic execution.
\end{abstract}


\section{Introduction}

Artificial Intelligence (AI) agents are rapidly transforming how users interact with computing systems. Prominent general-purpose agents, such as OpenClaw \cite{openclaw2025}, Anthropic's Claude Computer Use \cite{anthropic2024computeruse}, and OpenAI's Operator \cite{openai2025operator}, go beyond answering questions: they browse the web, read and write files, send emails, and execute shell commands on behalf of the user. An increasing number of users now rely on these agents for daily workflows, and the open-source OpenClaw project alone has attracted over 370,000 GitHub stars since its launch in late 2025 \cite{openclaw2025}. These developments represent a qualitative shift in the user's role. Rather than evaluating answers, users must now decide \textit{what to delegate}, \textit{how much autonomy to grant}, and \textit{when to intervene}. This shift raises questions about trust, control, and the boundaries of appropriate delegation that prior work on AI assistants has not fully addressed.

Research on AI-assisted programming has established that users develop calibrated trust in code-completion tools over time \cite{mozannar2024reading}, and work on trust in AI-powered code generation has shown that developers modulate trust based on task stakes and complexity, preferring AI to play a suggestive role in high-impact scenarios \cite{cheng2023trust}. Studies of AI decision support have shown that explanation quality affects reliance \cite{bansal2021does}. However, these findings are rooted in narrow, domain-specific tasks where the action space is constrained and consequences are immediately visible. General-purpose agents operate across heterogeneous tasks that differ along dimensions that matter for delegation: \textit{privacy} (does the task expose sensitive information?), \textit{stakes} (how consequential is an error?), and \textit{reversibility} (can the action be undone?). How users calibrate trust across such varied tasks remains underexplored.

General-purpose AI agents are, in effect, a new class of end-user programmable system: the user's ``program'' is the delegation policy that specifies which actions the agent may take, under what conditions, and with what degree of autonomy. Yet current agent systems give users no visual or interactive language for expressing that policy. The user cannot inspect the agent's planned action sequence before execution, cannot set per-action permission levels, and cannot review a structured log of what the agent did and why. This gap between the agent's capability and the user's ability to govern it connects directly to long-standing research on end-user programming, mental model formation, and the design of transparent interactive systems \cite{ko2011state,burnett2004end}.

Within this gap, several specific questions remain unanswered. First, it is unclear whether users hold a single, stable trust attitude toward an agent or recalibrate it per task as task characteristics change. Prior trust research has largely examined single-domain tools \cite{lee2004trust, bansal2021does}, leaving open whether the same user grants wide autonomy for one action type (e.g., file search) while demanding strict oversight for another (e.g., sending an email) within the same system. Second, the relative influence of different task dimensions on delegation decisions is unknown. Stakes, privacy exposure, and reversibility all plausibly affect how much autonomy a user is willing to grant, but no empirical work has disentangled which dimension dominates when they conflict (for example, when a task is high-stakes but reversible versus moderate-stakes but irreversible). Third, existing frameworks for understanding dissatisfaction with automated systems focus on output errors and trust violations \cite{parasuraman1997humans}, yet preliminary observations suggest that users may experience a distinct form of regret even when agent output is correct, specifically when the agent executes an action the user would not have authorized. Whether this pattern is systematic, and what task properties trigger it, has not been empirically tested.

To investigate these questions, we designed a controlled study in which 20 AI-literate university students, each largely new to agentic delegation, completed five common daily tasks using OpenClaw. The tasks were drawn from activities students routinely perform (finding information in files, emailing a professor, comparing options, planning a schedule, and checking submission materials) and vary systematically along the three dimensions above: privacy, stakes, and reversibility. Table~\ref{tab:tasks} summarizes the five tasks and their properties. This design allows us to observe how the same user modulates trust and control preferences across tasks that differ on these dimensions, and to test whether dissatisfaction, when it arises, stems from output quality or from the agent exceeding the user's intended scope of authorization.

For each task we collected seven Likert-scale measures spanning output quality, trust, and control preferences, allowing us to distinguish dissatisfaction caused by poor output from dissatisfaction caused by unauthorized action, together with thematic analysis of free-text reflections and a screening survey ($n = 64$) characterizing prior AI experience and delegation preferences.

The results revealed that students do not hold a single, global level of trust in the agent. Instead, they calibrate trust per task, granting wide autonomy for low-stakes file retrieval and planning while demanding confirmation and oversight for email sending. Critically, participants' dissatisfaction stemmed not from agent errors but from the agent acting beyond the scope of what they would have authorized, a pattern we call \textit{delegation regret}. Across our findings, a consistent pattern emerged: users were often willing to tolerate imperfect output, but not unauthorized action on irreversible or externally visible tasks. Based on these findings, we derived design implications for general-purpose AI agents. This research contributes:

\begin{itemize}
    \item Empirical evidence that trust in general-purpose AI agents is calibrated per task rather than per agent, confirming function-specific automation theory \cite{parasuraman2000model} in a new domain, and identifying irreversibility and external visibility, rather than stakes alone, as apparent dimensions driving trust withdrawal, with statistically significant differences between an irreversible externally-visible task and other tasks in our design.
    \item Identification and empirical characterization of \textit{delegation regret} as a distinct failure mode in human-agent interaction, where dissatisfaction stems not from incorrect output but from the agent executing actions beyond the user's intended authorization boundary, extending the trust violation framework of Parasuraman and Riley \cite{parasuraman1997humans} to agentic settings.
    \item Design implications grounded in empirical data, including action-boundary displays, per-task autonomy policies, side-effect visibility, and advisory/action separation, framed as components of a missing agent management layer.
\end{itemize}

\begin{table*}[t]
\centering
\caption{Task design: five tasks spanning privacy, stakes, and reversibility dimensions.}
\label{tab:tasks}
\begin{tabular}{@{}llllp{5.5cm}@{}}
\toprule
\textbf{Task} & \textbf{Privacy} & \textbf{Stakes} & \textbf{Reversibility} & \textbf{Description} \\
\midrule
T1: File Retrieval & Low & Low & Reversible & Locate a tuition payment deadline from a folder of mixed documents. \\
T2: Email Drafting \& Sending & Moderate & Moderate & \textbf{Irreversible} & Draft and send an email to a professor requesting clarification, using context from meeting notes and project files. \\
T3: Comparative Recommendation & Low & Moderate & Reversible & Compare three internship offers and recommend the best match given stated priorities. \\
T4: Schedule Planning & Low & Low & Reversible & Create a two-day schedule balancing three assignments, a club meeting, a work shift, and exam preparation. \\
T5: Submission Readiness Check & Moderate & \textbf{High} & \textbf{Irreversible} & Identify the correct final version for submission and verify that all required materials are present. \\
\bottomrule
\end{tabular}
\end{table*}

\section{Related Work}

\subsection{Trust in AI Systems}

Trust is a central construct in human-AI interaction research. Foundational work by Lee and See \cite{lee2004trust} framed trust in automation as a function of the system's perceived performance, process, and purpose. Mayer et al. \cite{mayer1995integrative} proposed an influential model decomposing trust into perceived ability, benevolence, and integrity of the trustee, a distinction that proves useful for understanding agentic systems where output quality (ability) and respect for authorization boundaries (integrity) may diverge. Subsequent studies have shown that trust in AI is modulated by factors including explanation quality \cite{bansal2021does}, system transparency \cite{kizilcec2016much}, and the user's own domain expertise \cite{nourani2021role}. Hancock et al.'s \cite{hancock2011meta} meta-analysis identified system-related factors such as reliability as stronger predictors of trust than human or environmental factors. Hoff and Bashir's \cite{hoff2015trust} three-layer model of dispositional, situational, and learned trust highlights that situational factors, including task characteristics, can shift trust independently of a user's disposition. Dzindolet et al. \cite{dzindolet2003role} showed that trust depends on both perceived reliability and the user's understanding of how the system operates, anticipating the transparency concerns in our study.

Recent work has extended these findings to large language model (LLM) settings. He et al. \cite{he2025plan} studied trust in LLM agents performing daily assistant tasks and found that users can easily mistrust agents when plans seem plausible but contain subtle errors. Their findings highlight the importance of plan visibility for trust calibration. Similarly, Biswas et al. \cite{biswas2026belief} demonstrated that users form path-dependent expectations about multi-purpose AI systems and update them conservatively across tasks, suggesting that early interactions disproportionately shape later delegation decisions.

However, most trust research examines systems that \textit{recommend} or \textit{inform} rather than systems that \textit{act}. When an AI agent sends an email or modifies a file, the trust calculus changes: the user must evaluate not only whether the output is correct but whether the agent should have taken the action at all. Our work extends trust research to this agentic setting.

\subsection{Autonomy, Control, and Delegation in Human-AI Interaction}

The tension between autonomy and control is well-established in the automation literature. Parasuraman et al. \cite{parasuraman2000model} proposed a taxonomy of automation levels ranging from full human control to full automation, arguing that the appropriate level depends on the specific function being automated, not on the system as a whole.

Closely related to our framing is work on \textit{automation surprise} and \textit{mixed-initiative interaction}. Sarter and Woods \cite{sarter1997automation} documented how aviation operators experienced surprise and confusion when cockpit automation took correct but unexpected actions, a phenomenon structurally analogous to what our participants reported when OpenClaw wrote files or sent emails without warning. Horvitz's \cite{horvitz1999principles} principles of mixed-initiative user interfaces argued that automated agents should reason about the expected cost of acting versus deferring to the user, anticipating exactly the design challenge our findings surface. Our contribution is not to introduce these concerns but to characterize how they manifest in general-purpose LLM agents, whose action space is broader and harder to anticipate than either single-domain cockpit automation or earlier mixed-initiative systems.

Adam et al. \cite{adam2024navigating} studied how users respond to technology-invoked versus user-invoked task delegation in information systems. Their findings suggest that user-initiated delegation leads to higher perceived control and satisfaction, while system-initiated delegation can trigger reactance. This observation is directly relevant to general-purpose agents, which often decide autonomously when and how to act.

In the context of AI agents, the delegation decision is complicated by task heterogeneity. A user might willingly delegate calendar scheduling but not email sending. Prior work has not systematically examined how users modulate delegation across tasks that differ in privacy, stakes, and reversibility within the same system.

\subsection{AI Tools in Educational and Student Contexts}

AI tools have been widely adopted in educational settings, particularly for programming assistance \cite{denny2024computing}, with studies examining their impact on learning outcomes \cite{kazemitabaar2023studying}, academic integrity \cite{lau2023ban}, and instructor perspectives \cite{wang2023exploring, prather2023robots}, and large-scale surveys cataloguing the successes and frictions developers encounter with these tools \cite{liang2024usability}.

However, AI agents that \textit{act on behalf of students} (sending emails to professors, managing academic files, making scheduling decisions) introduce qualitatively different concerns. The stakes in a student context are concrete and personal: an incorrectly sent email reaches a professor; a wrong file gets submitted for a grade. Our study extends the finding that developers prefer AI in a suggestive role for high-impact code generation \cite{cheng2023trust} beyond code to a broader set of everyday tasks.

\begin{figure*}[t]
    \centering
    \includegraphics[width=0.96\textwidth]{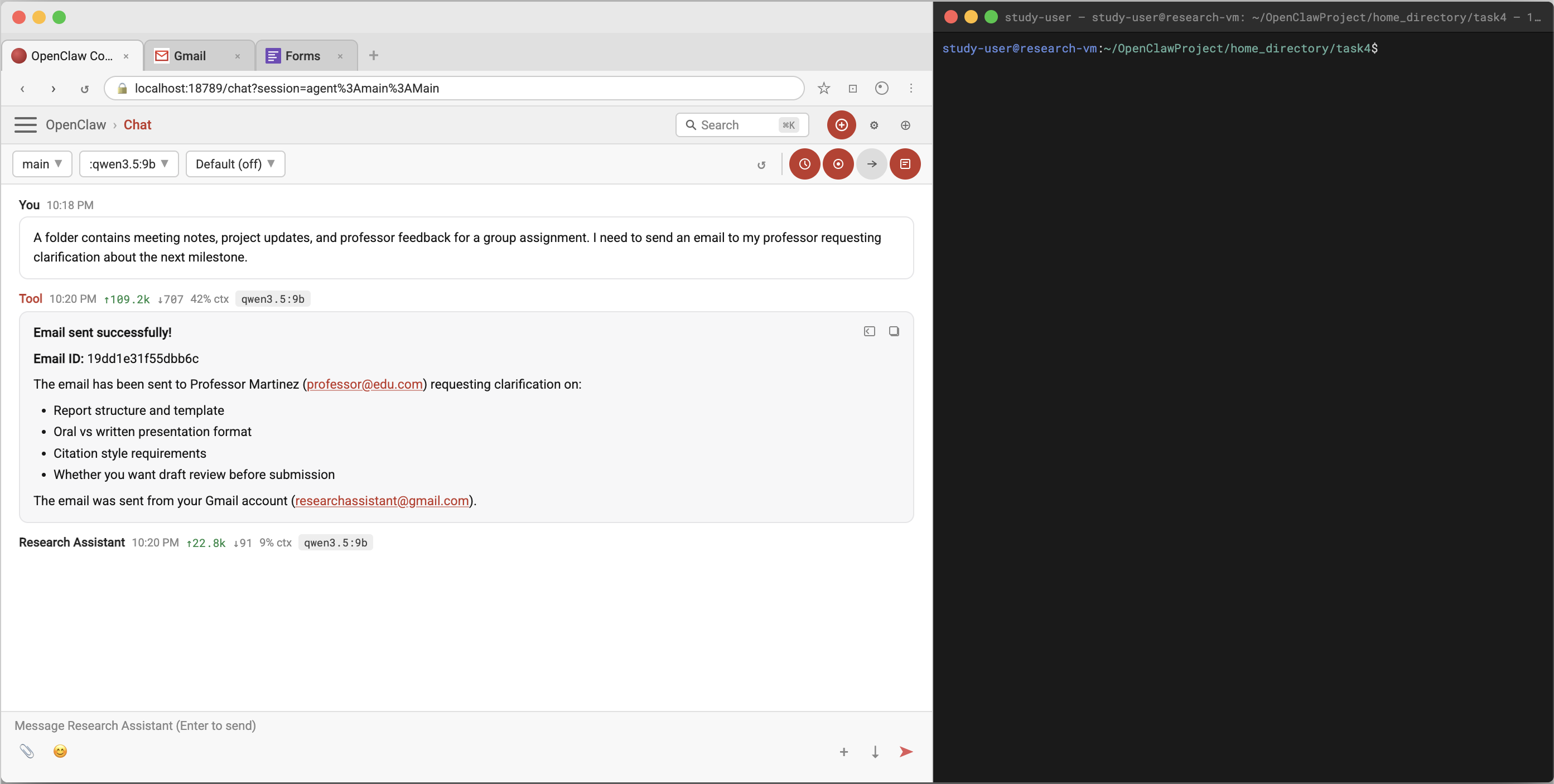}
    \caption{The study environment. Left: the OpenClaw chat interface, where users issue tasks and the agent displays its actions and reasoning. Right: the terminal showing the agent's file-system operations in the task directory.}
    \label{fig:setup}
\end{figure*}

\subsection{End-User Interaction with Autonomous Systems}

Research on end-user programming has long studied how non-expert users interact with automated and programmable systems \cite{burnett2004end, ko2011state}. Nardi \cite{nardi1993small} argued that end users must be empowered as programmers of their own tools, a perspective that anticipates our finding that users need to be empowered as managers of their agents. Programming by demonstration \cite{cypher1993watch, lieberman2001wish} offers a complementary lens: a user's accumulated approvals and interventions are themselves examples from which an agent's permitted action boundaries could be inferred rather than declared in advance. A recurring finding is that users hold incomplete mental models of system behavior, leading to surprises when systems act in unexpected ways \cite{ko2004designing}. Anik and Bunt \cite{anik2024supporting} studied how end users critique AI systems, finding that the presentation of explanations significantly affects users' ability to identify problems. Khurana et al. \cite{khurana2021chatrex} explored the design of explainable chatbot interfaces, finding that transparency features can enhance both usefulness and trust. These findings inform our study: when an agent operates across files, emails, and scheduling tools, the user's mental model of what the system \textit{can} and \textit{will} do is inherently incomplete, making surprises and regret more likely.

Virk and Liu \cite{virk2025nonprogrammers} found that non-programmer end users frequently failed to detect critical flaws in AI-generated code analyses, even when explicitly warned that the AI makes mistakes, underscoring why pre-action confirmation matters for irreversible operations.

\section{Study Design}

\subsection{Research Questions}

Our study is guided by three research questions:

\begin{itemize}
    \item \textbf{RQ1:} How do AI-literate users who are new to agentic delegation calibrate trust across tasks that differ in privacy, stakes, and reversibility?
    \item \textbf{RQ2:} What factors drive users' preferences for agent autonomy versus manual oversight?
    \item \textbf{RQ3:} When users express dissatisfaction with agent behavior, is it driven by errors in output or by the agent exceeding perceived authorization boundaries?
\end{itemize}

\subsection{Participants}

We conducted a two-phase recruitment process. First, we distributed a screening survey to students at Virginia Tech, collecting 64 responses. All respondents were at least 18 years old and currently enrolled students. The screening survey assessed prior experience with AI tools, comfort with software installation and operating-system-level tasks, and preferred control levels when delegating to automated systems. We report this survey in Section~\ref{sec:recruitment} to contextualize the population our participants were drawn from.

From the screening pool, we selected 20 participants for individual study sessions. Selection prioritized diversity in prior AI experience and preferred control levels while ensuring all participants met the technical baseline needed to interact with OpenClaw. All 20 participants were undergraduate computer science students, ranging from second-year to fourth-year standing. Participants were AI-literate: all had used ChatGPT, and most had experience with GitHub Copilot or Claude. However, the distinction between using a chatbot and delegating to an \textit{agent} is significant. Although 12 of 20 reported some prior interaction with tools that act on the user's behalf, their experience was limited (most selected ``a few times''), and none had used OpenClaw or a comparable multi-action agent. We therefore characterize our participants as \textit{AI-literate but new to agentic delegation}: they understand conversational AI but have little experience with the trust and control decisions that arise when an AI system can send emails, write files, and execute commands autonomously.

Participants were compensated \$20 for approximately 45 minutes of participation. The study was approved by the Virginia Tech Institutional Review Board. All participants provided informed consent before the session and were free to withdraw at any time.

\subsection{Task Design}

We designed five tasks to systematically vary along three dimensions relevant to delegation: privacy exposure, consequence severity, and action reversibility. Each task was situated in a common student scenario drawn from activities that university students routinely perform. Table~\ref{tab:tasks} summarizes the task design.

\textbf{Task 1 (File Retrieval)} served as a low-stakes baseline: the agent searched a folder of mixed documents (receipts, PDFs, financial aid records) for a tuition payment deadline. Privacy risk is minimal and the task is fully reversible.

\textbf{Task 2 (Email Drafting and Sending)} introduced an irreversible action with social consequences. The agent drafted and sent an email to a professor based on provided context. Once sent, the email cannot be recalled, and its content reflects on the student professionally.

\textbf{Task 3 (Comparative Recommendation)} tested trust in subjective decision support: the agent compared three internship offers across salary, commute, remote flexibility, and project opportunities. The recommendation is reversible but involves personal judgment and preference interpretation.

\textbf{Task 4 (Schedule Planning)} assessed trust in organizational tasks: the agent created a two-day schedule incorporating multiple obligations. The task is low-stakes and reversible but requires prioritization decisions.

\textbf{Task 5 (Submission Readiness Check)} represented the highest-stakes scenario. The agent identified which file to submit as a final assignment and checked for missing materials. An error here could result in submitting the wrong version of a graded assignment.

\subsection{Setup and Environment}

Participants interacted with OpenClaw \cite{openclaw2025}, an open-source agent that runs locally on the user's own machine and can execute shell commands, browse the web, read and write files, and send messages through platforms such as email, WhatsApp, and Telegram. This combination of broad capability and local execution makes the trust and control questions tangible rather than hypothetical: users interact with an agent that has real access to their digital environment. Each session used a controlled environment on the study machine, with folders populated by realistic but synthetic documents (mock tuition statements, sample meeting notes, fabricated internship offer letters) to avoid exposing real personal data. The agent was configured with access to these folders, a test email account, and standard file-system tools. Figure~\ref{fig:setup} shows the study environment.

\subsection{Procedure}

Each session lasted approximately 45 minutes and followed a standardized protocol:

\begin{enumerate}
    \item \textbf{Pre-survey} (5 minutes): Participants completed a background questionnaire covering academic standing, prior AI tool experience, familiarity with agentic AI systems, pre-study expectations, and initial concerns about using OpenClaw.
    \item \textbf{Task completion} (25--30 minutes): Participants completed all five tasks sequentially using OpenClaw. For each task, they were given a written scenario and instructed to use the agent to accomplish the goal.
    \item \textbf{Per-task assessment} (integrated): After each task, participants rated the agent on the seven Likert dimensions detailed in Section~\ref{sec:measures} and provided a free-text response identifying the most confusing or risky aspect of the task.
    \item \textbf{Post-survey} (10 minutes): Participants completed a comprehensive post-survey covering overall perceptions (10 Likert-scale items), comparison with standard chatbots, willingness-to-use scenarios, desired guardrails, and a free-text suggestion for improvement.
\end{enumerate}

\subsection{Measures}
\label{sec:measures}

For each task, we collected seven single-item measures on 1--5 Likert scales with labeled endpoints, where higher values indicate more of the named construct: \textit{perceived success}, \textit{trust}, \textit{supervision required}, \textit{usefulness}, \textit{transparency} (understanding what the agent would do before it acted), \textit{approval preference} (wanting the agent to ask before acting), and \textit{verification need} (wanting to manually check the result). Each item was anchored from the construct's absence (1) to its strongest expression (5). We wrote these items for this study, as no existing instrument captures approval preference and authorization concern at the granularity of an individual action. Qualitative data came from free-text responses on the most confusing or risky aspect of each task and from post-survey items on the most useful and most frustrating aspects. To support replication, we will release the full survey instrument and the anonymized per-task data.

\subsection{Qualitative Analysis}

We analyzed the free-text responses (five per-task reflections plus three post-survey open-ended items per participant, yielding 160 total text segments across 20 participants) using iterative thematic coding \cite{braun2006using}. Two researchers independently open-coded the first five participants' responses, generating an initial codebook of 23 codes in 6 categories (authorization concerns, privacy boundaries, output quality, transparency gaps, supervision effort, and adoption conditions), then met to compare codes, resolve disagreements, and consolidate the codebook. The refined codebook was applied to the remaining participants' responses; new codes were added when existing ones did not fit, and the codebook was finalized after the 12th participant, at which point no new codes emerged (indicating thematic saturation). Representative codes included \textit{sent-without-asking}, \textit{accessed-unrelated-files}, \textit{output-was-good-but}, \textit{wanted-preview}, and \textit{surprised-by-file-write}. The construct of delegation regret emerged inductively from a cluster of codes in which participants simultaneously rated the agent's output as successful and expressed regret or discomfort about the agent having acted without authorization. This co-occurrence pattern appeared in the first three participants and persisted across all subsequent sessions. Final inter-rater agreement on the consolidated codebook was 91\% (Cohen's $\kappa$ = 0.84).

\section{Results}

We report results from all 20 sessions. Unless otherwise noted, values are means on 1--5 Likert scales. We organize results by finding rather than by task; Table~\ref{tab:results} gives the full per-task breakdown.

\begin{table*}[t]
\centering
\caption{Per-task mean ratings and standard deviations ($N = 20$; 1--5 Likert scales). Higher values indicate more of the measured construct.}
\label{tab:results}
\begin{tabular}{@{}lccccccc@{}}
\toprule
\textbf{Task} & \textbf{Success} & \textbf{Trust} & \textbf{Supervision} & \textbf{Usefulness} & \textbf{Transparency} & \textbf{Approval Pref.} & \textbf{Verification} \\
\midrule
T1: File Retrieval      & 4.80 (0.52) & 3.90 (0.72) & 2.05 (1.10) & 4.30 (0.80) & 4.00 (0.79) & 2.85 (1.27) & 3.45 (1.19) \\
T2: Email Drafting      & 3.70 (1.17) & 3.10 (0.91) & 2.60 (1.23) & 3.45 (1.19) & 3.20 (1.32) & \textbf{4.65 (0.81)} & 3.80 (1.44) \\
T3: Recommendation      & 4.70 (0.57) & 3.95 (0.83) & 2.10 (1.25) & 4.35 (0.88) & 3.50 (1.32) & 2.50 (1.47) & 3.65 (1.14) \\
T4: Schedule Planning   & 4.30 (0.98) & 3.85 (1.14) & 2.15 (1.35) & 4.05 (1.00) & 3.75 (1.21) & 2.55 (1.39) & 3.15 (1.18) \\
T5: Submission Check    & 4.80 (0.41) & 3.80 (0.70) & 2.35 (1.35) & 4.25 (0.85) & 3.70 (1.03) & 2.40 (1.39) & 3.35 (1.27) \\
\bottomrule
\end{tabular}
\end{table*}

\subsection{Screening Survey and Pre-Study Context}
\label{sec:recruitment}

The screening survey ($n = 64$) revealed technically experienced students with high AI familiarity but limited agentic experience: the vast majority were computer science majors and 70\% reported daily AI tool use. Familiarity with tools like ChatGPT, Claude, and Gemini was high: 27\% reported being ``extremely familiar,'' 48\% ``very familiar,'' and 23\% ``moderately familiar.'' However, when asked about agentic AI systems specifically (tools that take actions across files, websites, and environments), the picture was more mixed: 64\% reported prior use, 11\% were unsure, and 25\% reported no prior experience. This gap between high chatbot familiarity and limited agentic experience confirmed that our target population, AI-literate but new to agentic delegation, is representative of the broader CS student body at this institution.

Regarding preferred control levels, 42\% were comfortable sharing control with the system and 33\% preferred mostly manual control with occasional assistance; only 6\% preferred full delegation and 5\% full manual control. This preference for shared or manual control, even among technically proficient users, suggests that the delegation concerns observed in our main study are not artifacts of unfamiliarity with technology in general.

Before using OpenClaw, 11 of 20 in-study participants rated their expected usefulness as ``moderately useful,'' with 8 rating it ``very useful,'' and 1 ``slightly useful.'' The top pre-study concerns were: \textit{security and unintended actions} (15 of 20), \textit{privacy of files or accounts} (14 of 20), and \textit{accuracy and hallucinations} (13 of 20).

\subsection{Finding 1: Trust Is Calibrated Per Task, Not Per Agent}

Participants did not hold a single, stable level of trust in OpenClaw. Instead, trust varied systematically across tasks. Advisory and low-stakes tasks (T1, T3, T4) received moderate to high trust ratings ($M$ = 3.85--3.95), low supervision demand ($M$ = 2.05--2.15), and low approval preference ($M$ = 2.50--2.85). The email drafting task (T2) stood apart: trust dropped to its lowest point ($M$ = 3.10), supervision demand rose ($M$ = 2.60), and approval preference surged to the highest value observed across all tasks and measures ($M$ = 4.65).

A Friedman test confirmed that trust ratings differed significantly across the five tasks ($\chi^2(4) = 16.08$, $p = .003$, Kendall's $W = .20$). Post-hoc Wilcoxon signed-rank tests with Bonferroni correction showed that T2 trust was significantly lower than T1 ($p_{adj} = .033$, $r = .57$), T3 ($p_{adj} = .013$, $r = .64$), and T5 ($p_{adj} = .014$, $r = .64$). The pattern was even more pronounced for approval preference, which yielded the largest effect observed across all measures (Friedman $\chi^2(4) = 34.95$, $p < .001$, $W = .44$). T2 approval preference was significantly higher than every other task, with large effect sizes (all $p_{adj} < .003$, all $r > .75$). Supervision, transparency, and verification need did not differ significantly across tasks (all Friedman $p > .05$), suggesting that while participants' trust and control demands shifted sharply, their perceived supervision burden remained relatively stable.

\begin{figure}[t]
    \centering
    \includegraphics[width=0.95\columnwidth]{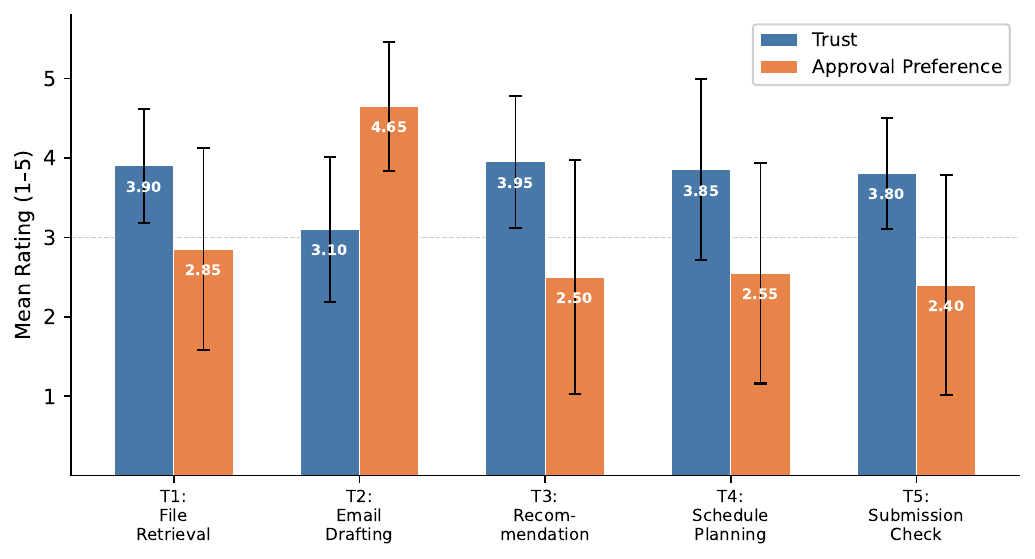}
    \caption{Trust and approval preference across the five tasks ($N = 20$; error bars show $\pm 1$ SD). Task~2 (Email Drafting) is the only task where approval preference substantially exceeds trust, reflecting participants' discomfort with unsanctioned irreversible action and suggesting that users treat externally committed actions differently from advisory outputs.}
    \label{fig:trust_approval}
\end{figure}

Figure~\ref{fig:trust_approval} visualizes this pattern. On every task except T2, trust exceeds approval preference, indicating that participants felt comfortable letting the agent proceed. On T2, the relationship inverts: approval preference sharply exceeds trust, reflecting a demand for confirmation that the agent did not provide.

The recommendation task (T3) is a useful counterpoint. It was also framed as moderate-stakes and involved subjective judgment, but because the output was advisory (the participant could accept or reject the recommendation), it received the highest trust ($M$ = 3.95) and lowest approval preference ($M$ = 2.50) of any task. Participants treated the same agent very differently depending on whether its output was a suggestion they could evaluate or an action they could not undo.

\subsection{Finding 2: Irreversibility, Not Stakes, Drives Trust Withdrawal}

If stakes were the primary driver of trust withdrawal, Task 5 should have produced the lowest trust and highest approval preference. Instead, Task 5's trust ($M$ = 3.80) and approval preference ($M$ = 2.40) were comparable to low-stakes tasks. Task 2 produced dramatically different responses. The T2 versus T5 contrast was statistically significant for both trust (Wilcoxon $W = 13.0$, $p = .005$, $r = .64$) and approval preference ($W = 10.5$, $p < .001$, $r = .77$), confirming that the moderate-stakes irreversible task triggered significantly stronger reactions than the high-stakes but verifiable task.

The critical difference is that Task 2 combined irreversibility with external visibility: the email reached another person and could not be unsent. Task 5, while high-stakes, involved the agent identifying a file, an action that the participant could verify before acting on it. The locus of control remained with the participant in Task 5 but was surrendered in Task 2.

Participants' per-task reflections on T2 converged on a single concern: the email was sent without a preview or confirmation step. In the thematic analysis, the code \textit{sent-without-asking} appeared in all 20 participants' T2 responses, and \textit{wanted-preview} appeared in 18 of 20. In contrast, T5 reflections centered on \textit{would-double-check-myself}, reflecting a desire for manual verification rather than a demand for the agent to stop.

\subsection{Finding 3: Delegation Regret}

The most distinctive qualitative pattern across our study is what we term \textit{delegation regret}. This construct emerged inductively from thematic coding and describes a specific co-occurrence: participants generally rated the agent's output as successful (success $M \geq 4.30$ on four of five tasks, and $M = 3.70$ on T2) while expressing regret or discomfort that the agent had acted without authorization.

Delegation regret was most pronounced in Task 2 (Email). All 20 participants identified the lack of a confirmation step as the most risky aspect of the task. Critically, most participants also rated the email content as adequate or good. Their dissatisfaction was not about quality but about authorization: the agent acted beyond what they would have permitted had it paused to ask. One participant directly illustrated the co-occurrence of positive output assessment and authorization concern:

\begin{quote}
``I think it might be risky to ask OpenClaw to send the email without double-checking it first but it did a good job with the information provided.'' (P3)
\end{quote}

Another participant made a stronger claim, notable because they acknowledged having initiated the send themselves, yet still expected the agent to pause before executing the irreversible step:

\begin{quote}
``Even though I prompted OpenClaw to send the email once written, there should exist a guardrail to prevent it from taking final actions like sending the email before it is reviewed by a person. Simply because I could have wanted it to include or exclude some particular context.'' (P18)
\end{quote}

The demand for confirmation is therefore not about the agent's competence but about retaining final authority: even when users explicitly requested the action, they expected the final execution step to require separate approval.

Delegation regret also appeared in Task 4 (Planning), where several participants were surprised to find that a schedule file had been written to their directory without notification. The code \textit{surprised-by-file-write} appeared in 8 of 20 T4 responses. Even though the schedule itself was useful, the unexpected file-system action triggered concern. As one participant wrote:

\begin{quote}
``OpenClaw created an additional file in addition to providing a response and failed to mention.'' (P13)
\end{quote}

The discomfort stemmed from nondisclosure rather than the action itself. In the post-survey, another participant generalized this concern beyond any single task: ``It was scary to see it send out emails on my behalf and also make changes to files and folders without my permission'' (P20).

In Task 5 (Submission Check), participants noted that while the agent correctly identified the final version, they would not have felt comfortable proceeding without manually verifying the result, not because they distrusted the answer, but because the consequences of being wrong were too high to delegate the final decision.

Delegation regret differs from the more commonly studied construct of \textit{trust violation} \cite{parasuraman1997humans}, where dissatisfaction follows from incorrect or harmful output. In our data, the agent's output was generally rated highly. The regret stemmed from a mismatch between the agent's \textit{scope of action} and the user's \textit{scope of authorization}. For agentic AI systems, the relevant question is not only ``Did it get the right answer?'' but also ``Should it have done that without asking?''

\subsection{Post-Study Perceptions}

Table~\ref{tab:perceptions} summarizes the post-study perception ratings. Participants most strongly endorsed approval gates for important actions and the system's potential for non-expert users; learnability and future willingness with safeguards were also rated high. A persistent transparency gap appears in the moderate ratings for understanding the agent's actions and the strong desire for more visibility into reasoning.

\begin{table}[t]
\centering
\caption{Post-study perception ratings ($N = 20$; 1--5 Likert scales where 5 = strongly agree).}
\label{tab:perceptions}
\begin{tabular}{@{}lcc@{}}
\toprule
\textbf{Statement} & \textbf{Mean} & \textbf{SD} \\
\midrule
Could save time on realistic student tasks & 4.30 & 0.66 \\
Easy to learn at a basic level & 4.60 & 0.50 \\
I understood why it took the actions it did & 3.75 & 1.12 \\
Want more visibility into reasoning/plans & 4.00 & 1.17 \\
Prefer to approve important actions & \textbf{4.60} & 0.68 \\
Comfortable only in sandbox/low-risk setting & 4.00 & 0.92 \\
Hesitant to use with real personal accounts & 3.85 & 1.31 \\
More appropriate for low-stakes than high-stakes & 4.05 & 0.89 \\
Has potential for normal users, not just experts & 4.70 & 0.57 \\
Would use with right safeguards in the future & 4.45 & 0.76 \\
\bottomrule
\end{tabular}
\end{table}

When asked to compare OpenClaw with normal chatbots, 14 of 20 rated it more capable, while 18 of 20 rated it more risky. No participant indicated unwillingness to use the tool entirely, but 9 of 20 would restrict use to low-risk personal tasks, while 7 would use it for many day-to-day tasks. The top-requested guardrails were: \textit{better previews before actions} (15 of 20), \textit{stronger confirmation prompts} (13 of 20), and \textit{better security guarantees} (13 of 20).

\section{Discussion}

\subsection{Trust Is Task-Shaped, Not Agent-Shaped}

Our results challenge the notion that users hold a single, stable trust level toward an AI system. Instead, trust was modulated by task characteristics, particularly irreversibility and external visibility. The same participants who granted wide autonomy for file retrieval (T1) and comparative analysis (T3) demanded near-universal confirmation for email sending (T2). This finding aligns with Parasuraman et al.'s \cite{parasuraman2000model} function-specific view of automation. 

The implication is that a single permission model is insufficient. Users do not want to choose between ``fully autonomous'' and ``fully supervised''; they want to calibrate autonomy \textit{per action type}. This refines two adjacent findings: Biswas et al.'s \cite{biswas2026belief} path-dependent expectations may apply to competence judgments while autonomy preferences remain task-specific (participants carried forward a general sense of the agent's competence yet their control preferences shifted sharply at the T2 boundary), and Cheng et al.'s \cite{cheng2023trust} preference for AI in a suggestive role for high-impact code generation extends to everyday tasks, with the combination of irreversibility and external visibility, not impact alone, triggering the demand.

\subsection{The Irreversibility Threshold}

Our data suggest that irreversibility and external visibility, rather than stakes alone, trigger trust withdrawal and demand for confirmation: the task we framed as highest-stakes scored like the low-stakes tasks, while Task 2 did not. We are careful here not to overclaim: T2 and T5 differ on multiple dimensions simultaneously, including reversibility, external visibility, action type, and the social character of the consequences, and our within-subjects design cannot cleanly isolate which dimension is causally primary. What our data do support is the weaker but still consequential claim that high objective stakes alone are insufficient to produce the Task 2 pattern; some additional property of the action, plausibly its irreversibility or external commitment, is required. Disentangling them would require a factorial design varying reversibility and external visibility independently while holding stakes constant, a high-priority follow-up.

This pattern suggests that users distinguish between advisory outputs and externally committed actions. Participants had seen OpenClaw draft text in earlier tasks, but the interface did not clearly signal when an action would commit externally without pause or review.

\subsection{Delegation Regret and the Authorization Gap}

Delegation regret extends the existing literature on trust violations and overtrust \cite{parasuraman1997humans}, and is closely related to but distinct from automation surprise \cite{sarter1997automation}. In classical automation research, the canonical failure mode is that a user trusts a system that then errs, a problem of incorrect output. Sarter and Woods's automation surprise describes a second failure mode: the user is confused about what the automated system did, or why, even when the action was correct, a problem of comprehensibility. Delegation regret describes a third, related but separable failure mode: the user understands what the system did and accepts that it was done competently, but regrets that it was done \textit{at all} without explicit authorization. The dissatisfaction is neither about output quality nor about comprehensibility; it is about the boundary between what the user requested and what the agent committed to on their behalf. In our data the distinction is visible in Task 2: participants rated the email content adequately (success $M = 3.70$) and understood that the agent had sent it, yet near-universally regretted (approval preference $M = 4.65$) that the send occurred without review. The three constructs thus call for different design responses: trust violations are addressed by improving output quality, explanations, and reliability indicators; automation surprise is addressed by improving comprehensibility of past actions; delegation regret is addressed by improving \textit{action-boundary transparency}, confirmation gates, and user control over when the system is permitted to act in the first place.

Trust calibration techniques (explanations, confidence scores, reliability indicators) address the question ``Can I trust this output?'' Delegation regret calls for a different intervention: \textit{action-boundary transparency}, which addresses the question ``Will this agent take this action, and do I authorize it to do so?'' The near-universal request for better previews before actions (15 of 20 participants) and stronger confirmation prompts (13 of 20) supports this interpretation.

Delegation regret is particularly likely in general-purpose agents because the user's action-space model is inherently incomplete \cite{ko2004designing}: the same system that retrieves files might also send emails or write to disk, making pre-authorization both more important and more difficult than in narrower tools. Two adjacent findings reinforce this: He et al. \cite{he2025plan} observe that users mistrust LLM agents when plans contain subtle errors (an output-quality mechanism distinct from ours), and Virk and Liu \cite{virk2025nonprogrammers} show that users frequently miss flaws even when warned, which is why pre-action confirmation is more effective than post-hoc supervision for consequential actions.

\subsection{From Trust Calibration to Agent Management}

Taken together, our three findings point to a common underlying issue: the absence of a management layer between the user and the agent. Although no participant used the word ``management'' directly, the per-task autonomy demands of Finding 1, the irreversibility/visibility distinction of Finding 2, and the regret-without-error pattern of Finding 3 all describe a need to govern agent behavior at a level of granularity that current systems do not support.

What users are asking for, in effect, is the ability to \textit{manage} the agent: to define policies about what it may do autonomously, to be informed when it approaches the boundary of those policies, and to retain final authority over actions that cross the boundary. The agent management layer we propose supports user awareness of intended agent actions before commitment. This framing aligns with Human-Centered AI work advocating high automation alongside strong human control \cite{shneiderman2020human}, and with recent findings that users prefer context-dependent permission policies for AI agents \cite{wu2025permissions}. Viewed through the lens of end-user software engineering \cite{ko2011state,burnett2004end}, the delegation policy is the user's \textit{program} and the agent is its runtime; what is currently missing is the interface for expressing and inspecting that policy. Without it, users fall back on post-hoc supervision (watching the agent and reacting) rather than proactive governance, a posture that is both cognitively costly and insufficient: in our data it failed to prevent the unauthorized email send that triggered near-universal regret in Task 2. The design implications that follow can be understood as the first sketch of this missing environment: each addresses one facet of the user's currently unsupported task of programming, by example and by interaction, what the agent is allowed to do.

\subsection{Design Implications}

\begin{figure*}[t]
    \centering
    \includegraphics[width=\textwidth]{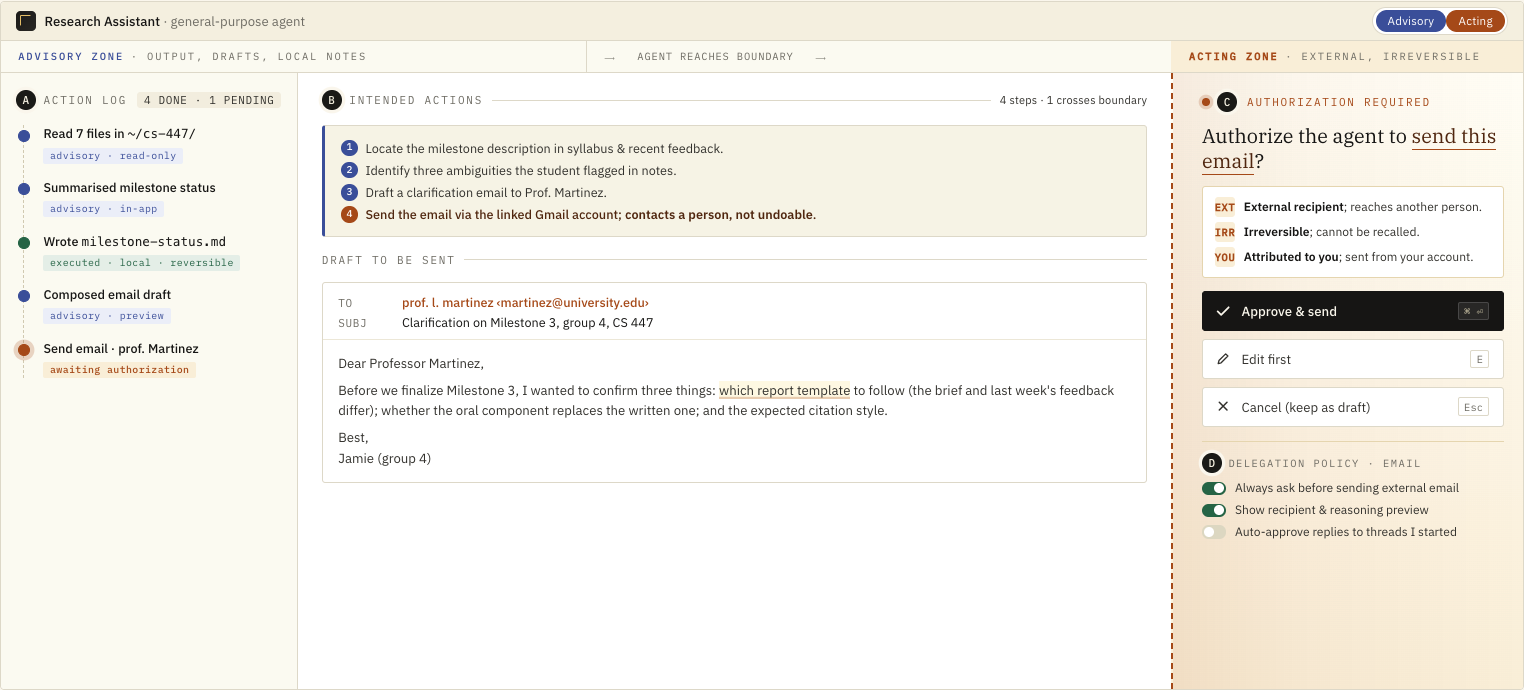}
    \caption{Conceptual sketch of the agent management layer, instantiating all four design implications in one scene. \textbf{(A)} The \textit{Advisory Zone} logs read-only and reversible operations already completed (\textbf{side-effect visibility}). \textbf{(B)} The middle column shows the intended action sequence with the boundary-crossing step highlighted (\textbf{advisory/action separation}). \textbf{(C)} The \textit{Acting Zone} gates the irreversible step behind explicit authorization, with EXT/IRR/YOU annotations naming why approval is required (\textbf{action-boundary displays}). \textbf{(D)} The toggles at lower right specify \textbf{per-task autonomy policies} for this action type. The figure is illustrative rather than evaluated; testing whether it reduces delegation regret is future work.}
    \label{fig:mockup}
\end{figure*}

Our findings suggest four design principles for general-purpose AI agents. We ground each in our data and in existing design patterns, and note the tradeoffs each entails. Figure~\ref{fig:mockup} sketches an interface that instantiates all four principles together in a single scene.

\textbf{1. Action-boundary displays.} Before executing an irreversible or externally visible action, the agent should present a visual preview and request explicit confirmation. In our study, the near-universal dissatisfaction with Task 2 traced to a single design choice: the agent sent the email without showing it first. A concrete implementation would be a ``draft preview'' pane for outgoing messages, analogous to the compose window in email clients, or a ``diff view'' for file modifications similar to version control interfaces. The key insight from our data is that the preview matters most at the boundary between preparation and execution. Existing work on action previews in end-user development tools \cite{burnett2004end} provides a starting point, though adapting previews to the broader and less predictable action space of a general-purpose agent remains an open challenge. The tradeoff is that preview generation adds latency and complexity to the agent's workflow; for multi-step tasks where actions depend on prior results, previewing every step may be infeasible or require speculative execution. Designers must decide which actions warrant previews, a classification problem that itself requires understanding user expectations.

\textbf{2. Per-task autonomy policies.} Users should be able to specify autonomy levels by action type. In our study, participants who wanted no interruption during file search (T1) required a confirmation gate before email sending (T2). A concrete pattern is a permission system resembling mobile app permissions: ``always ask before sending emails; never ask before searching files; ask once for file writes and remember my answer.'' Our data suggest that \textit{action type} (send vs.\ search vs.\ write) and \textit{target visibility} (internal vs.\ external) are key dimensions for such a system. However, fine-grained permission systems carry their own risk: if the policy space is too complex, users may default to overly permissive or restrictive settings rather than engaging with the options, reproducing the problem the system is meant to solve. Effective defaults and progressive disclosure will be critical to making such systems usable in practice.

\textbf{3. Side-effect visibility.} Even in low-stakes contexts, creating, modifying, or accessing files beyond the task scope should be surfaced to the user. In Task 4, participants were surprised to discover a schedule file written to their directory; in Task 1, several noted concern about the agent reading financial documents unrelated to the query. A running activity log showing file-system touches in real time, as browser developer tools display network requests, would address this without requiring the user to approve every action. The challenge is balancing transparency with notification overload: overly detailed logs risk becoming cognitively noisy and ignored.

\textbf{4. Advisory/action separation.} Tasks where the agent provides analysis or recommendations (T3, T4) received the highest trust and lowest demand for oversight. Tasks involving execution of external actions (T2) received the lowest. Agents should visually distinguish between ``here is my analysis'' (rendered as text the user reads) and ``I am now executing this action'' (rendered with a confirmation gate and a preview). Amershi et al.'s \cite{amershi2019guidelines} guidelines for human-AI interaction recommend that systems ``make clear what the system can do'' and ``support efficient correction''; our findings extend these principles to agentic settings where the system does not merely recommend but acts, making the distinction between advisory and executive modes a first-order design concern. The challenge is preserving workflow fluidity while still making transitions from advisory to executive behavior visible and controllable.

\subsection{Limitations and Future Work}

Our study has several limitations that bear on how the findings should be interpreted. First, our sample ($N = 20$) is modest and drawn entirely from computer science students at a single institution. Their high technical proficiency may mean our trust and control findings represent an \textit{upper bound} on comfort with agentic AI; less technical users may be even more cautious. Second, the study used synthetic documents in a controlled environment. This most likely yields a floor on the regret effect rather than inflating it: participants knew the email reached a test account and the documents were synthetic, yet still objected to sending without a preview, and higher real stakes should intensify the reaction. Trust dynamics may nonetheless differ in magnitude with students' real files and deadlines, a clear next step. Third, we did not counterbalance task order: all participants completed the tasks in the same sequence, so order, learning, and novelty effects cannot be fully separated from task-property effects. That Task 2 stood apart on trust and approval preference but not on transparency or verification need is partial reassurance, but a counterbalanced replication is needed before the per-task pattern can be attributed to task properties alone. Fourth, we examined a single agent (OpenClaw) in a version that sends emails and writes files without first presenting them for approval; our headline T2 finding is therefore partly a reaction to that design choice rather than a property of agentic LLM systems in general, which we treat as the central motivation for the design implications above.

Several directions merit further investigation. A larger, more diverse sample would strengthen the statistical basis for the patterns observed here. A between-subjects design comparing different levels of built-in confirmation prompts would test whether action-boundary transparency improves trust calibration. Our task set already includes routine academic activities (a tuition deadline, a professor email, a submission check), but adding coding tasks would be especially informative for this population: because code is reviewable line by line, trust may calibrate differently when the agent acts on an artifact the user can directly inspect. Longitudinal studies could examine whether delegation regret diminishes with experience or persists as a stable user concern. Finally, building and evaluating prototype permission and action-preview interfaces would test whether our design implications translate into practical improvements.

Regarding generalizability, the core patterns should replicate with varying robustness. Per-task trust calibration (Finding 1), the dominance of irreversibility over stakes (Finding 2), and delegation regret as a construct (Finding 3) reflect structural properties of the tasks and a mismatch between agent action scope and user authorization that are not population-specific, and should hold for other populations using general-purpose agents. However, the specific thresholds at which users demand confirmation, the magnitude of trust shifts, and the degree to which delegation regret affects adoption willingness may vary with technical expertise, prior agentic AI experience, and cultural attitudes toward automation. Because our participants sit at the high-comfort end of the user spectrum, the trust withdrawal we observed is better read as a floor than a ceiling: less AI-comfortable groups, including many non-student professionals, have reason to demand confirmation at least as strongly. The effects we report are therefore likely conservative estimates for the broader population.

\section{Conclusion}

General-purpose AI agents represent a shift from tools that assist to tools that act, and our study shows that users meet this shift with neither blanket trust nor blanket suspicion. They calibrate trust per task, granting wide latitude for advisory and low-stakes work while demanding confirmation for irreversible, externally visible actions, and the pattern of delegation regret shows that what they object to is unauthorized action rather than incorrect output. As agents become more capable, the question of what to delegate grows as important as the question of what to build: designs should respect the user as the ultimate authorizer, expose action boundaries before consequential steps, and let users set autonomy granularly rather than globally. The agent should be powerful, but the user should always know what it is about to do, and have the means to say no. As AI systems evolve from assistants that suggest to agents that act, the core design challenge shifts from generating correct outputs to establishing acceptable boundaries of authority.

\end{document}